\documentclass[10pt,conference]{IEEEtran}

\usepackage[english]{babel}
\usepackage[utf8]{inputenc}
\usepackage{url}
\usepackage{enumitem}
\usepackage{microtype}
\usepackage{todonotes}
\usepackage{listings}
\usepackage{graphicx}

\IEEEoverridecommandlockouts \IEEEpubid{\makebox[\columnwidth]{\copyright~2018 IEEE \hfill} \hspace{\columnsep}\makebox[\columnwidth]{ }}

\begin{document}

\title{Trustworthy Configuration Management for Networked Devices using Distributed Ledgers
\thanks{This work has been supported by the German Federal Ministry of Education
and Research, project DecADe, grant 16KIS0538 and the German-French
Academy for the Industry of the Future.
\protect\\
\indent Author's version -- Final paper presented at 2018 IEEE/IFIP International Workshop on Decentralized Orchestration and Management of Distributed Heterogeneous Things (DOMINOS) co-located with the Network Operations and Management Symposium (NOMS)~\cite{179826}.}
}

\author{\IEEEauthorblockN{Holger Kinkelin, Valentin Hauner, Heiko Niedermayer and Georg Carle}
\IEEEauthorblockA{Technische Universität München, Department of Informatics\\
Chair of Network Architectures and Services \\
85748 Garching b. München, Germany\\
\{lastname\}@net.in.tum.de}}

\maketitle

\begin{abstract}

Numerous IoT applications, like building automation or process control of industrial sites, exist today.
These applications inherently have a strong connection to the physical world.
Hence, IT security threats cannot only cause problems like data leaks but also safety issues which might harm people.

Attacks on IT systems are not only performed by outside attackers but also insiders like administrators.
For this reason, we present ongoing work on a \emph{configuration management system~(CMS)} that provides control over administrators, restrains their rights, and enforces separation of concerns.
We reach this goal by conducting a configuration management process that requires \emph{multi-party authorization} for critical configurations to achieve Byzantine fault tolerance against attacks and faults by administrators.
Only after a configuration has been authorized by multiple experts, it is applied to the targeted devices.
For the whole configuration management process, our CMS guarantees \emph{accountability} and \emph{traceability}.
Lastly, our system is \emph{tamper-resistant} as we leverage \emph{Hyperledger Fabric}, which provides a distributed execution environment for our CMS and a blockchain-based distributed ledger that we use to store the configurations.
A beneficial side effect of this approach is that our CMS is also suitable to manage configurations for infrastructure shared across different organizations that do not need to trust each other.

\end{abstract}

\IEEEpeerreviewmaketitle

\section{Introduction}
\label{intro}

The \emph{Internet of Things (IoT)} connects devices from small sensors and actuators to large machines.
Home or building automation, machine monitoring, and process control of industrial plants are only some IoT application examples~\cite{iotreport}.

One inherent property of IoT is a strong connection to the physical world.
For this reason, security weaknesses can result in privacy problems when sensitive data is leaked or persons get injured if safety mechanisms fail.
Consequently, IoT systems require a high IT security standard whose steady maintenance is challenging in the age of \emph{Advanced Persistent Threats (APT)}.
APTs often target system administrators directly by attacks like spear phishing or waterholing~\cite[p.~37/38]{symanticsecreport}.
Consequences for the whole IoT system are severe, as the attacker can abuse the conquered administrative rights.

Another important scenario are administrators that turned evil and who now abuse their rights, e.g., to steal company secrets or to harm their employer.
Such attacks can be categorized as insider attacks, which caused 10\%-32\% of data breaches in 2015 according to different studies~\cite[p.~53]{symanticsecreport}.
Both scenarios stress the need to defend against attacks that involve abused administrative rights~\cite[p.~10]{googlesecreport}.
However, this goal is hard to achieve, as administrative rights also typically give the opportunity to tamper or disable many traditional security solutions, such as security and incident event handling (SIEM), access control, or logging/auditing systems.

\textbf{Contributions:}
In this paper, we present ongoing work on a \emph{configuration management system (CMS)} that provides control over system administrators, restrains their rights, and helps to enforce separation of concerns.
We reach this goal by conducting a configuration management process that requires \emph{multi-party authorization (MPA)} for critical configurations to achieve Byzantine fault tolerance against attacks and faults by administrators.
Only after a configuration has been reviewed and authorized by a set of independent experts, the managed devices retrieve the configuration from our CMS and apply it locally.
The different parties that need to authorize a configuration can be specified on a per-device basis, making it possible to take into account the criticality of a device. For the whole configuration management process of our CMS, we guarantee \emph{accountability} and \emph{traceability}.
Lastly, our system is \emph{tamper-resistant} as we leverage \emph{Hyperledger Fabric}, which provides a distributed execution environment for our CMS and a blockchain-based distributed ledger to store configurations.
A beneficial side effect of this approach is that our CMS is also suitable to manage configurations for infrastructure shared across different organizations which only share a limited amount of trust.

\textbf{Structure:}
Background and related work are explained in Section~\ref{ch:bgrw}.
We define requirements in Sect.~\ref{ch:req}.
The design of the CMS is explained in Sect.~\ref{ch:des} and an outlook on the ongoing implementation is given in Sect.~\ref{ch:impl}.
Sect.~\ref{ch:discussion} discusses intermediate results before we conclude in Sect.~\ref{ch:conclusion}.

\section{Background and Related Work}
\label{ch:bgrw}

\subsection{Configuration Management Systems (CMS)}
\label{sec:ansi}

Literature neither defines the term \emph{CMS} precisely nor specifies which features a CMS has exactly.
Commonly, any system that standardizes and facilitates the way how devices are configured is regarded as a CMS.

A core concept of a CMS is to describe configurations in a serialized, structured representation that can be applied automatically to devices using an appropriate tool~\cite[Sect. 2]{conftoolsurvey}.
This concept is known as \emph{Infrastructure as Code (IaC)}~\cite{morris2016} and its most simple instantiation would be a configuration shell script.
However, in recent years, more elaborate toolsets were created.
Examples include Chef, Puppet and \emph{Ansible}~\cite{ansible}, which denotes a serialized configuration as a \emph{playbook}.

Besides expressing and applying configurations, a CMS can have further tasks like enforcing a workflow that, for instance, includes reviewing configurations~\cite[Sect. 2.3.6]{conftoolsurvey}.
For this reason, a CMS can be seen as an intermediary between administrators and managed devices.
However, most CMSs are lacking workflow enforcement today~\cite[Sect. 3.3.6]{conftoolsurvey}.

\subsection{Blockchain-Based Distributed Ledger Technology (DLT)}
\label{sec:bc}

A \emph{distributed ledger} is \textit{``a type of database that is spread across multiple sites''}.
\textit{``Records are stored one after the other in a continuous ledger''} and \textit{``can only be added when the par\-ti\-ci\-pants reach a quorum''}~\cite[p. 17]{dlt}.
Resulting properties are \emph{non-modifiability} and \emph{non-erasability} of data and that participants of the ledger are not required to fully trust each other.
One example of distributed ledger implementations are \emph{blockchains}.
Use cases for DLT include, for instance, the \emph{Bitcoin} payment system, whose ledger is public, but also enterprise applications, whose ledgers are typically private.

\emph{Hyperledger Fabric}~\cite{hyperledgerpro, hyperledgerdoc} is an enterprise blockchain.
The business logic of an application running on top of the Fabric network is written in \emph{chaincode}.
Each chaincode is installed on a set of \emph{endorsing peers}, together with a dedicated \emph{endorsement policy} that specifies which peers have to \emph{endorse} any \emph{transaction} on the chaincode.
A transaction creates or modifies a data record stored in the ledger, which consists of a blockchain and a state database.
To modify the ledger, a client \emph{invokes} some operation of the chaincode by sending a \emph{transaction proposal} to the endorsing peers that have the chaincode installed.
Each of them executes the operation without modifying the ledger and returns the signed execution result in the form of an \emph{endorsement} to the client.
If all results match, the client will send the transaction together with the set of endorsements to the \emph{ordering service}, a composite of several independent nodes that gathers transactions from all clients and eventually puts them into a block that is delivered to all peers in the network.
Each peer receiving the block, now in the role of a \emph{committing peer}, appends the whole block to its blockchain copy. Afterwards, it checks the validity of each transaction in the block, i.e., if the endorsement policy is fulfilled and the execution results of the different peers match.
If this holds, the peer will apply the transaction's execution result to its copy of the state database.

\section{Requirements}
\label{ch:req}

This section defines essential requirements on our CMS.

\textbf{R1 -- Multi-party authorization:}
Critical configurations must be reviewed and authorized by multiple experts.

\textbf{R2 -- Accountability and traceability:}
The CMS must log all steps of the configuration management process to provide the means to understand who configured what in which way.

\textbf{R3 -- Tamper-resistance:}
The CMS must be resistant against attacks that even abuse administrative rights. This includes protected execution of the configuration management process and non-erasable and non-forgeable storage of configuration data.

\section{Design}
\label{ch:des}

This section provides an overview of our system and showcases how functional components interact.

\subsection{Approach}
\label{sec:approach}

Our approach is based on the idea to manage \emph{configuration requests (CR)} in a \emph{configuration management system (CMS)}.
A CR is a request targeted to one or several managed devices to apply the configuration included in the CR.
As a \emph{configuration}, we understand any serialized representation of a system state of a managed device.
In contrast to many existing CMSs, our system never applies a CR directly to a target device.
Instead, a CR must undergo a validation process before it is applied.
We furthermore assume that remote shell access is disabled on devices and that a \emph{configuration daemon}, a trustworthy and automated application, runs on devices which guarantees that configurations are applied once they are validated.

\begin{figure}[h]
  \centering
    \includegraphics[width=\columnwidth]{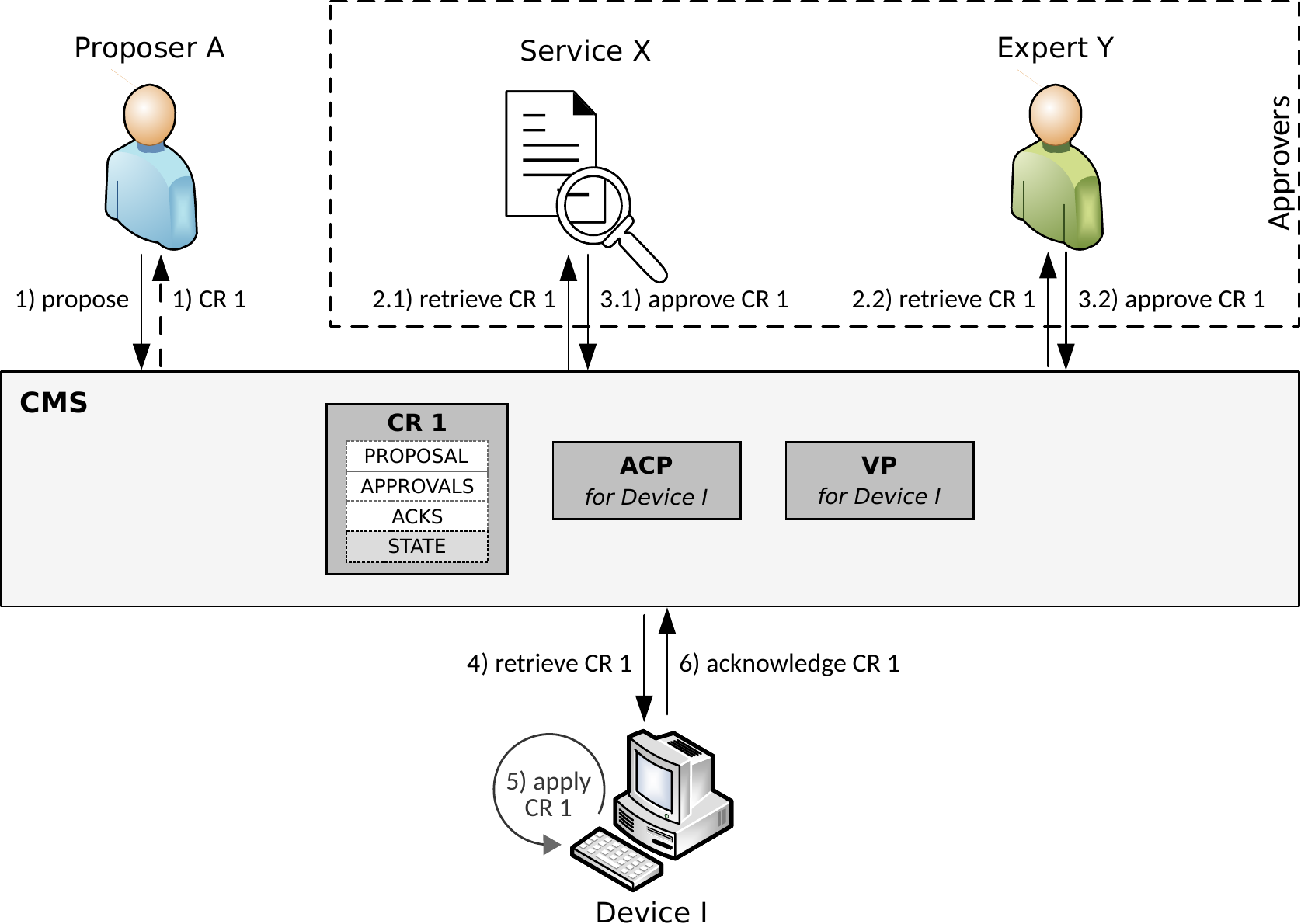}
  \caption{System architecture and interactions between stakeholders}
  \label{overview}
\end{figure}

A CR is a composite data structure consisting of a \emph{proposal}, a set of \emph{approvals} and a set of \emph{acknowledgements}.
The proposal includes the actual configuration and the set of target devices.
The approvals indicate who has approved the CR, while the acknowledgements indicate which devices have applied it.
In order to guarantee authenticity, authorization and accountability, proposal, approvals and acknowledgements must be digitally signed by the corresponding actor.
Furthermore, each CR has a life cycle, represented by the states \emph{proposed}, \emph{valid} and \emph{acknowledged}.
The current state is directly stored in the CR and updated by the CMS.

A CR starts its life in the \emph{proposed} state. An administrator, acting here in the role of a \emph{proposer}, \emph{proposes} the desired configuration together with the set of target devices to the CMS, c.f. step~1 in figure~\ref{overview}.

The CMS first checks the permissions of the proposer according to the \emph{access control policy (ACP)}.
After the access has been granted, a CR is created and stored in the CMS, making it accessible for the subsequent validation process.

To reflect a device's criticality, this process can be more or less complex.
This means that a CR for highly critical devices must undergo a more thorough validation to reach the \emph{valid} state than a CR for less critical devices.
The individual tests and other conditions of the validation process are specified in the corresponding \emph{validity policy (VP)} stored in the CMS.

The individual tests are performed by entities we call \emph{approvers}.
Approvers can either be computer programs running on dedicated machines or human experts, e.g. other system administrators.
Tests may include simple syntax checks which can automatically be performed by programs or security audits performed by human experts.
In order to minimize the risk of individual fraudulent approvers or mistakes, the VP can stipulate that tests need to be repeated by $n$ approvers. In case test results differ from another, a majority vote or other rules, like $m$ out of $n$ approvals, can be applied.

As a next step, an approver has to \emph{retrieve} the desired CR, c.f. step~2. After having performed a test successfully, she \emph{approves} the CR, c.f. step~3. This may include a reference into the VP to express which test has been performed together with the test result.
Once the CMS has received the approver's submission, a new approval is appended to the CR.

By time, approvers conduct more and more tests stipulated by the VP, which results in numerous approvals stored as part of the respective CR.
As soon as the VP is fulfilled for a CR, the CMS updates the CR's state to \emph{valid}.
If a device notices a new valid CR targeted to itself, it \emph{retrieves} this CR and automatically applies the configuration, c.f. step~4 and 5.

After having successfully applied the configuration, the device \emph{acknowledges} the respective CR, c.f. step~6.
As soon as every target device of the CR has applied and acknowledged the configuration, the CR reaches the \emph{acknowledged} state.

\subsection{3-Tier Architecture}
\label{ch:3ta}

\begin{figure}[h]
  \centering
    \includegraphics[width=0.9\columnwidth]{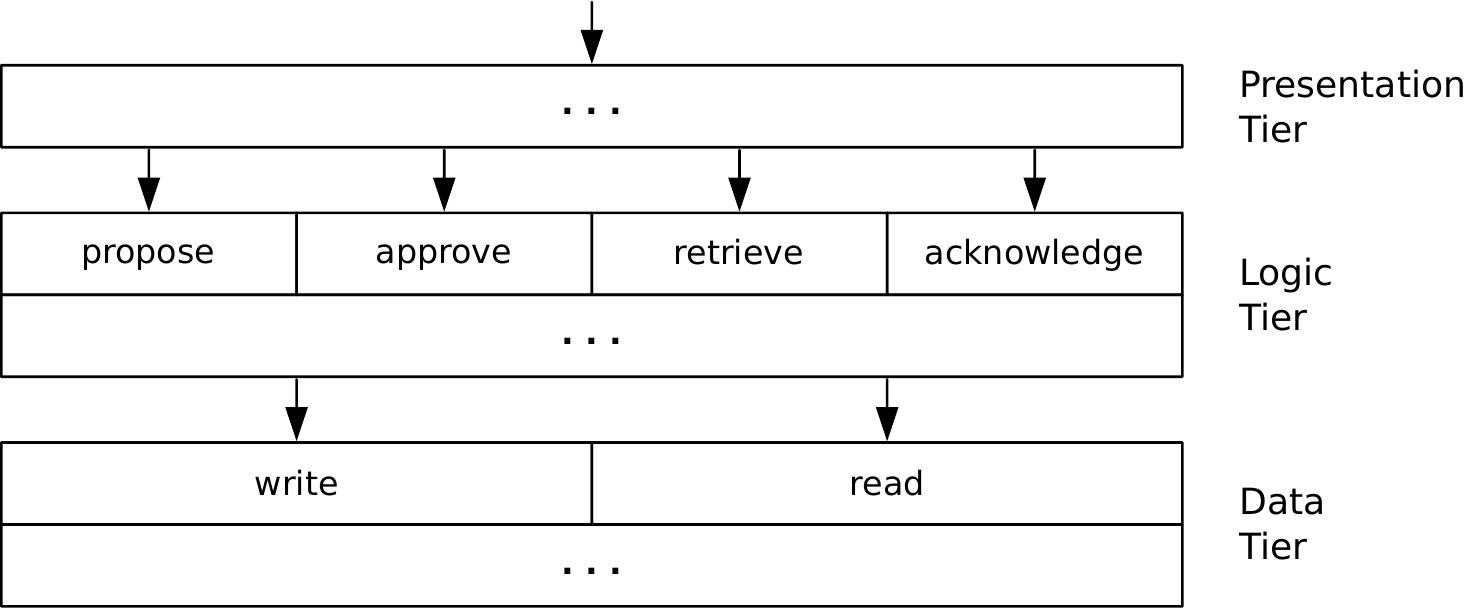}
  \caption{3-tier architecture}
  \label{tiers}
\end{figure}

Our design follows the 3-tier architectural pattern, c.f. figure~\ref{tiers}.
The topmost \emph{presentation tier} serves as a user interface.
The \emph{logic tier} provides the business logic of our CMS and offers an API with operations to \emph{propose} or \emph{approve} a CR, but also to \emph{retrieve} or \emph{acknowledge} one, as described in Sect.~\ref{sec:approach}.
The \emph{data tier} serves as the interface to the used data storage. It offers \emph{write} and \emph{read} operations to the logic tier.

\section{Implementation Concept}
\label{ch:impl}

This section provides an outlook on the ongoing implementation of our CMS.
The presentation tier is implemented as a command line interface (CLI) and runs on the client.
The essential role in our implementation, however, is played by \emph{Hyperledger Fabric}, c.f. Sect.~\ref{sec:bc}.
The logic as well as the data tier run in a Fabric network, composed of a set of peers and an ordering service.
The whole business logic described in Sect.~\ref{sec:approach} is executed as chaincode on these peers.
The data tier corresponds to the ledger managed by the peers, consisting of the blockchain and the state database.
Consequently, we utilize Hyperledger Fabric in two ways: as a distributed execution environment and as a distributed storage.

\begin{figure}[h]
  \centering
    \includegraphics[width=\columnwidth]{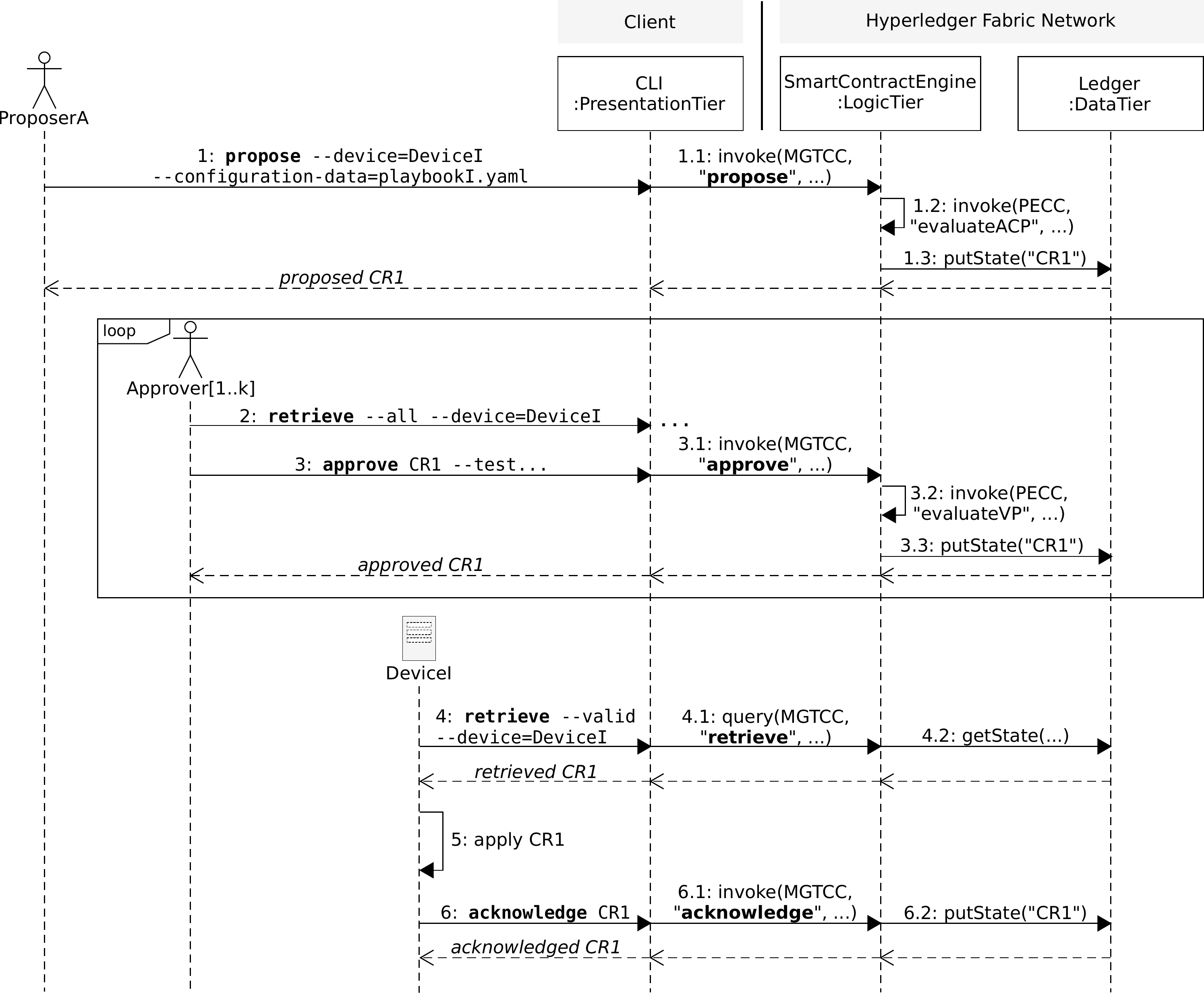}
  \caption{Message flow through tiers}
  \label{flow}
\end{figure}

The sequence diagram in figure~\ref{flow} shows how the different tiers work together in our implementation.
To run our application, one has to set up and start a Fabric network first.
The chaincodes that handle the business logic have to be installed on the endorsing peers.
Our implementation includes two chaincodes: the \emph{management chaincode (MGTCC)} used to \texttt{propose}, \texttt{approve}, \texttt{retrieve} and \texttt{acknowledge} a CR, and the \emph{policy evaluation chaincode (PECC)} used for evaluating if a CR fulfills an ACP and a VP, respectively.
The endorsement policies for the chaincodes depend on the use case: If, for instance, the system manages CRs for devices shared across different organizations, the endorsement policies may specify that a transaction has to be endorsed by peers of each organization in order to prevent the peers of one organization from acting maliciously.

A proposer proposes a CR via the CLI, c.f. step 1 in figure~\ref{flow}.
The command's arguments are the target devices and the configuration data itself, the latter in the form of an Ansible playbook, c.f. Sect.~\ref{sec:ansi}.
The CLI interprets the command and calls the logic tier by invoking the \texttt{propose} operation of the MGTCC (1.1).
The MGTCC first checks if the proposer is permitted to propose a CR for the specified devices according to the corresponding ACP.
For this, it calls the \emph{evaluateACP} operation of the PECC (1.2).
If the PECC returns a positive result, a new CR will be created and put into the ledger (1.3).

The chaincode invocation in step 1.1 leads to a transaction flow in the Fabric network that is not shown in the figure, but has been described in Sect.~\ref{sec:bc}.
Here, the proposer serves as the client and sends a transaction proposal to the endorsing peers, which will execute the \texttt{propose} operation and return their respective endorsement to the client.
Afterwards, the client sends the transaction together with the endorsements to the ordering service, which will eventually broadcast a new block to all peers of the network to be appended to their blockchain copies.
The peers check the transaction's validity and, if fulfilled, finally write the CR contained in the transaction to their copy of the state database.
Subsequently, the proposer is notified that the CR has been successfully proposed.

After the approvers have noticed a proposed CR, either by manually retrieving it from the system (2) or by getting notified dynamically, they review the CR.
If one of them agrees to the configuration, she will call the CLI to approve the CR (3).
The CLI again interprets the command and hands it over to the logic tier to invoke the \texttt{approve} operation of the MGTCC (3.1).
The MGTCC adds the new approval to the set of approvals contained in the CR.
Then, it calls the \emph{evaluateVP} operation of the PECC to check if the CR is already valid according to the corresponding VP (3.2).
If the PECC returns a positive result, the MGTCC will change the CR's status to \emph{valid}.
Then, it puts the modified CR into the ledger (3.3).
The transaction flow in the Fabric network is analogous to that in step 1.1.
Eventually, the approver will be notified that the approval succeeded.

The configuration daemon running on a device retrieves all valid CRs targeted to it (4).
Since retrieving a CR does not modify it, it is sufficient in step 4.1 to perform a \emph{query} using the \texttt{retrieve} operation of the MGTCC that actually gets the CR from the ledger (4.2).
This time, the transaction flow just involves multiple endorsing peers which execute the operation and return the signed results to the daemon, which verifies if all results are identical.

After having successfully retrieved a valid CR, the configuration daemon applies the configuration locally (5) without having to re-validate the CR as the valid state cannot be forged.
In particular, the playbook stored in the CR is now run using Ansible. In the end, the configuration daemon acknowledges the CR (6), followed by the call of the logic tier (6.1).
The \texttt{acknowledge} operation of the MGTCC adds a new acknowledgement to the CR's set of acknowledgements.

After each of the target devices has acknowledged the CR, the MGTCC changes the CR's state to \emph{acknowledged}. Then, it puts the modified CR into the ledger~(6.2).

\section{Discussion}
\label{ch:discussion}

As this paper describes work in progress, we have not yet conducted a full evaluation.
However, we want to discuss and compare our status quo with requirements defined in Sect.~\ref{ch:req}.

Our CMS fulfills the concept of \emph{multi-party authorization (MPA)} by requiring the agreement of multiple experts on a configuration contained in a CR, corresponding to R1.

The whole business logic is executed as chaincode in a Fabric network. Therefore, we leverage a Fabric network as a distributed execution environment and as a distributed storage, providing \emph{accountability}, \emph{traceability} and \emph{tamper-resistance} for all operations of the CMS, corresponding to R2 and R3.

The peer-to-peer structure of the network even allows to manage CRs for infrastructure shared across competing organizations that do not fully trust each other.
Each stakeholder can easily contribute its own nodes to the network and prevent other stakeholders from acting maliciously, e.g. by manipulating the result of a chaincode invocation.

\section{Conclusion}
\label{ch:conclusion}

In this paper, we argued that uncontrolled administrative access rights can pose a serious threat to the security of IoT systems or other networked systems.
We proposed a \emph{configuration management system (CMS)} that acts as an intermediate authority between administrators and managed devices.
The CMS is able to conduct \emph{multi-party authorization (MPA)} to achieve Byzantine fault tolerance against hazardous or faulty configurations.
To guarantee \emph{accountability}, \emph{traceability} and \emph{tamper-resistance}, we employ \emph{Hyperledger Fabric} as a distributed execution environment and as a distributed storage for the whole system and its data objects.
Furthermore, our CMS is suitable to manage configurations for infrastructure shared across different organizations that do not need to fully trust each other.
Future work includes, among others, finalizing and carefully evaluating the prototype implementation and examining how situations that require rapid responses can be handled in our CMS.

\section*{Acknowledgments}

The authors want to thank Hendrik Leppelsack, Marcel von Maltitz and Miguel Pardal for valuable input on the paper.
Our work has been supported by the German Federal Ministry of Education and Research (grant 01LY1217C, project DecADe).

\bibliographystyle{IEEEtran}
\bibliography{ref}

\end{document}